\documentclass[%
 reprint,
 amsmath,amssymb,
 aps,
pra,
]{revtex4-1}

\usepackage{graphicx}% Include figure files
\usepackage{dcolumn}% Align table columns on decimal point
\usepackage{bm}% bold math
\usepackage{float}
\usepackage[%Uncomment any one of the following lines to test 
text={7in,9.5in},centering,
]{geometry}

\begin{document}
\renewcommand{\sfdefault}{phv}
\preprint{APS/123-QED}

\title{Density Functional Theory 
Based on the \\ Electron Distribution on the Energy Coordinate}% Force line breaks with \\

\author{Hideaki Takahashi*}%
\affiliation{%
 Department of Chemistry, Graduate School of Science, Tohoku University,
 Sendai, Miyagi 980-8578, Japan 
}%

\date{\today}% It is always \today, today,

\begin{abstract}
We introduced a new electron density $n(\epsilon)$ by projecting the spatial electron density $n(\bm{r})$ onto the energy coordinate $\epsilon$ defined with the external potential $\upsilon(\bm{r})$ of interest. Then, a density functional theory (DFT) was formulated, where $n(\epsilon)$ serves as a fundamental variable for the electronic energy. It was demonstrated that the Kohn-Sham equation can also be adapted to the DFT that employs the density $n(\epsilon)$ as an argument to the exchange energy functional. An important attribute of the energy density is that it involves the spatially non-local population of the spin-adapted density $n(\bm{r})$ at the bond dissociation. By taking advantage of this property we developed a prototype of the static correlation functional employing no empirical parameters, which realized a reasonable dissociation curve for $\text{H}_2$ molecule.  

\begin{description}
\item[PACS numbers]
\pacs{} 31.10.+z, 31.15.E-
\end{description}
\end{abstract}
\maketitle

Kohn-Sham density-functional theory (KS-DFT) \cite{rf:kohn1965pr} has been successfully utilized to study various properties of molecule, cluster, and solid. It offers a versatile framework for describing the 
electronic correlation energies of matters  merely in terms of  electron density $n$. It, however, suffers from 
critical  problems, referred to as ``static correlation error(SCE)" and ``self-interaction error(SIE)", that emerge in the dissociations of chemical bonds even in  simplest molecules, e.g. $\text{H}_2$ or $\text{H}_2^+$. As reviewed in Ref. \cite{rf:cohen2008sci}, the former gives rise to serious destabilization in the electronic energy of a system with fractional spins at atomic sites, and it also leads to qualitative failures in nearly degenerate electronic states. The latter manifests itself as erroneous stabilization for a system with fractional charges, and consequently, tends to  delocalize  electrons over the system artificially. Importantly, it is diagnosed that the problematic situations arising in the applications of KS-DFT have their origins in  these specific errors \cite{rf:cohen2008sci}. Thus, eliminating the root of errors will extend the frontier of  DFT. \\
\indent  The source of the failures lies not in the formalism of DFT, but in the fundamental framework of the approximate exchange-correlation functional $E_{xc}[n]$ which carries the complex effects of electron repulsions. The Kohn-Sham equation for the non-interacting electrons involves the local potential defined as $\upsilon_{xc}[n](\bm{r})=\delta E_{xc}[n]/\delta n(\bm{r})$ at position $\bm{r}$. \textit{In principle}, it is possible to incorporate the \textit{non}-local information of the electron density in the construction of $\upsilon_{xc}(\bm{r})$. In practice, however,  due to the local density approximation (LDA) on which the basic framework of the approximate functionals is founded, only local or semilocal properties at $\bm{r}$ ($n(\bm{r}), \nabla n(\bm{r})$, and $\nabla^2 n(\bm{r}),\ldots$) are  utilized to yield the potential $\upsilon_{xc}(\bm{r})$. Here, we briefly illustrate the SCE which arises in describing a bond dissociation with LDA.  Suppose  that the chemical bond of $\text{H}_2$  is stretched, the exchange hole associated with a reference point $\bm{r}_\text{ref}$ on an atomic site will be split over the two separated atoms. Unfortunately, it is  unlikely that the local quantities at $\bm{r}_\text{ref}$  can take into consideration the other exchange hole shifted on the distal site. Thus, the hole depth at $\bm{r}_\text{ref}$  is to be estimated as half of the density $n(\bm{r}_\text{ref})$, which causes the erroneously high dissociation limit of $\text{H}_2$ within the LDA-based approach \cite{rf:becke2003jcp}. The underlying mechanism for the SIE can also be interpreted in terms of the defect arising from LDA. A new framework of approximation based on  a different starting point is, therefore, needed to overcome these problems. \\
\indent  We propose in this Letter to introduce a new distribution of electrons which serves as a fundamental variable in DFT. Explicitly, we project the electron density $n_{\sigma}(\bm{r})$ $(\sigma=\alpha \text{ or }  \beta)$ onto an energy coordinate $\epsilon$ to yield a new distribution, thus, 
\begin{equation}
n^e_\sigma[\upsilon_{\text{def}}](\epsilon)=\int d\bm{r} \delta (\epsilon - \upsilon_{\text{def}} (\bm{r}))n_{\sigma} (\bm{r}) 
\label{eq:energy_dist}
\end{equation}
where $\upsilon_{\text{def}}$ is the potential function introduced to define energy coordinate $\epsilon$ for electrons. Usually, the external potential $\upsilon_{\text{ext}}$ of interest is taken as the potential $\upsilon_{\text{def}}$. Hereafter, the dependence of the density $n^e_{\sigma}(\epsilon)$ on the potential $\upsilon_{\text{def}}$ is not explicitly indicated for the sake of brevity. We note  Eq. (\ref{eq:energy_dist}) is parallel in form to the definition of the solvent distribution function utilized in a novel theory of solutions in energy representation developed by Matubayasi et al \cite{rf:matubayasi2000jcp}. Their approach has been successfully applied to the calculation of solvation free energies of various solutes in solutions. As demonstrated below a rigorous framework of electronic DFT can also be founded  in terms of the energy distribution $n_{\sigma}^{e}(\epsilon)$ of Eq. (\ref{eq:energy_dist}), and further, the theory can be adapted to Kohn-Sham procedure provided some explicit functional $E_{xc}^{\upsilon}[n_{\sigma}^{e}]$ is given. 

\begin{figure}
\scalebox{0.48}[0.48]{\includegraphics{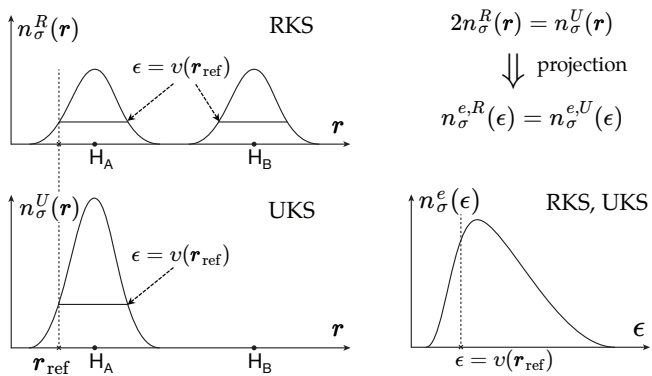}} % Here is how to import EPS art
\caption{\label{Fig1:epsart} Electron densities $n_{\sigma}^{\text{R}}$ (upper left) and $n_{\sigma}^{\text{U}}$ (lower left) of dissociated $\text{H}_2$ molecules, respectively, for the spin-restricted and unrestricted Kohn-Sham DFT. Lower right figure shows the energy distribution  $n_{\sigma}^{e}(\epsilon)$ given by the projections of $n_{\sigma}^{\text{R}}$ and $n_{\sigma}^{\text{U}}$. }
\end{figure}
\indent   The projection of the spatial distribution $n_{\sigma}$ onto the energy coordinate $\epsilon$ of one dimension (Eq.(\ref{eq:energy_dist})) smears the details of the density used to build the exchange hole function.  However,  as investigated by Parr and Berk in Ref. \cite{rf:parr1981plenum}, the electron densities in molecules as well as in atoms have contours nearly parallel to those of bare-nuclear Coulomb potentials. From the statistical point of view it seems quite natural to consider that the electron population is reasonably constant on the equi-energy surface of the external potential. Provided that this postulate is valid, the new type of DFT that is based on the energy electron density $n_{\sigma}^{e}(\epsilon)$ will not spoil the quality of the exchange functional. The advantage of introducing the distribution of Eq. (\ref{eq:energy_dist}), which is vital for the present development, is that the energy distribution can take into consideration the spatially non-local population of electrons. We illustrate in Fig.\ref{Fig1:epsart} the electron densities $n_{\sigma}^{\text{R}} (\bm{r})$ and $n_{\sigma}^{\text{U}} (\bm{r})$ yielded, respectively, with spin-restricted and unrestricted KS-DFT for a dissociated $\text{H}_\text{2}$ molecule. These distributions completely differ from each other in their spatial behaviors. The projections of these onto the energy coordinate $\epsilon$, however, provides exactly the same distribution $n_\sigma^e(\epsilon)$ when $2 n_{\sigma}^{\text{R}}(\bm{r})=n_{\sigma}^{\text{U}}(\bm{r}) $ is satisfied on site $\textsf{H}_\textsf{A}$. Thus, the symmetry-broken solution gives the same energy distribution as the symmetry-adapted density. Therefore, the value of the functional  $E_{xc}^{\upsilon}[n_{\sigma}^{e, \text{R}}]$ becomes in principle the same as  that of $E_{xc}^{\upsilon}[n_{\sigma}^{e, \text{U}}]$ for any choice of approximate functional $E_{xc}^{\upsilon}$ of the energy distribution. For SIE the energy distribution also shows a favorable behavior. Explicitly, the charge  delocalized over the dissociated  fragments has the same energy distribution as that for the charge distribution localized on a single site. Hence, the argument to the energy functional $E_{\text{ee}}[n^e]$ for the electron repulsion stays constant for the charge fragmentation. From the viewpoint of the functional development, this property is advantageous since the delocalized state has the same electron energy as the localized state.   \\
\indent We  now formulate the density functional theory where  the projected density $n^e(\epsilon)$ instead of $n(\bm{r})$ serves as a fundamental variable. First of all, it should be kept in mind that an energy distribution is constructed using  a  defining potential $\upsilon_{\text{def}}$ as shown in Eq.(\ref{eq:energy_dist}) and, hence, the information contents of an energy distribution differs from another one when they are associated with different $\upsilon_{\text{def}}$. For a given external potential $\hat{\upsilon}_{\text{ext}}$ the Hamiltonian $\hat{H}$ for $N$-electron system is given by
% hamiltonian
\begin{equation}
\hat{H}=\hat{T}+\hat{V}_{ee}+\hat{\upsilon}_{\text{ext}}
\label{eq:hamiltonian}
\end{equation}
where $\hat{T}$ and $\hat{V}_{ee}$ are, respectively, the kinetic energy and electron repulsion operators. 
The ground state energy $E_0$ of a system can be derived in the way parallel to the Levy's constraint search \cite{rf:levy1979pnas} for an $N$-electron system, thus, 
%Levy's constraint search
\begin{eqnarray}      \notag
E_{0} & = & \min_{\Psi\rightarrow N}\left\langle \Psi\right|\hat{H}\left|\Psi\right\rangle \\    \notag
 & = & \min_{n^{e}\rightarrow N}\left\{ \min_{\Psi\rightarrow n^{e}}\left\langle \Psi\right|\hat{H}\left|\Psi\right\rangle \right\} \\   \notag
 & = & \min_{n^{e}\rightarrow N}\left\{ \min_{\Psi\rightarrow n^{e}}\left\langle \Psi\right|\hat{T}+\hat{V}_{ee}\left|\Psi\right\rangle +\int d\epsilon\; n^{e}\left(\epsilon\right)\upsilon^{e}\left(\epsilon\right)\right\} \\    
 & = & \min_{n^{e}\rightarrow N}\left\{ F_{\upsilon}\left[n^{e}\right]+\int d\epsilon\; n_{}^{e}\left(\epsilon\right)\upsilon^{e}     \left(\epsilon\right)\right\} 
\label{eq: Levy}
 \end{eqnarray}\\
The first line in Eq. (\ref{eq: Levy}) expresses a minimization of the expectation value of $\hat{H}$ with respect to the antisymmetric wave function $\Psi$ with $N$ electrons. Below the second line, the energy is minimized in the space of $N$-rep. energy electron density $n^e$. In the third line the external potential $\upsilon^e$ is also represented on the energy coordinate defined by the potential $\upsilon_{\text{def}}$ in Eq. (\ref{eq:energy_dist}). Specifically, if $\upsilon_{\text{ext}}(\bm{r})$ itself is taken as $\upsilon_{\text{def}}(\bm{r})$, we have $\upsilon^e(\epsilon) = \epsilon$. In the fourth line we introduced a functional $F_{\upsilon}[n^e]$ which represents the sum of kinetic and electronic energies of $\Psi_{\text{min}}$ that yields the energy density $n^e(\epsilon)$. Importantly,  it is also possible to prove that  a one-to-one  correspondence establishes between the subset of $\upsilon$-rep. energy  densities $n^e$ associated with a potential $\upsilon_{\text{def}}$ and the subset of the potentials $\upsilon(\epsilon)$ defined on the same energy coordinate $\epsilon$. Thus, the Hohenberg-Kohn type variational principle \cite{rf:hohenberg1964pr}  holds for the energy distributions.  However, it should be noted that $F_{\upsilon}[n^e]$ is \textit{not} a universal functional of the argument $n^e$ since the energy distribution $n^e$ is specific to the choice of $\upsilon_{\text{def}}$. Later, we will demonstrate that some $\upsilon$-dependent exchange-correlation functional $E_{xc}^{\upsilon}[n^{e}]$ can actually be  constructed using the conventional DFT functionals as templates. The functionals for kinetic energy and electron-electron repulsion can be defined using the wave function $\Psi_{\text{min}}[n^e]$ which minimizes the expectation value of $\hat{T}+\hat{V}_{ee}$ as 
\begin{eqnarray}
T\left[n^{e}\right] & = & \left\langle \Psi_{\text{min}}\left|\hat{T}\right|\Psi_{\text{min}}\right\rangle 
\label{eq: kinetic}
\\
V_{ee}\left[n^{e}\right] & = & \left\langle \Psi_{\text{min}}\left|\hat{V}_{ee}\right|\Psi_{\text{min}}\right\rangle .
\label{eq: ee-repul}
\end{eqnarray}
\\
\indent Next, we consider the method to adapt the variational search of Eq. (\ref{eq: Levy}) to the Kohn-Sham procedure. Introducing a single determinant wave function $\Psi_{\text{SD}}$ of a non-interacting system we define a functional $G[n_e]$, thus, 
\begin{equation}
G\left[n^{e}\right]=\min_{n\rightarrow n^{e}}\left\{ \min_{\Psi_{\text{SD}}\rightarrow n}\left\langle \Psi_{\text{SD}}\right|\hat{T}\left|\Psi_{\text{SD}}\right\rangle +J\left[n\right]\right\}\;\;\; .
\label{eq: non_int}
\end{equation}
In Eq. (\ref{eq: non_int}), $J[n]$ is the Hartree energy of the electron density $n(\bm{r})$ which yields $n^e$.  Then, the kinetic energy $T_s[n^e]$ of the non-interacting system can be represented as 
\begin{equation}
T_{s}\left[n^{e}\right]=\left\langle \Psi_{\text{SD}}^{\min}\right|\hat{T}\left|\Psi_{\text{SD}}^{\text{min}}\right\rangle
\label{eq: kinetic_s}
\end{equation}
where $\Psi_{\text{SD}}^{\text{min}}$ is the wave function which minimizes $T+J$ in Eq. (\ref{eq: non_int}) under the constraint that its electron density gives $n^e$. We define the Hartree energy $J^{\prime}[n^e]$ as a functional of $n^e$
as
\begin{equation}
J^{\prime}\left[n^{e}\right]=\dfrac{1}{2}\int d\bm{r}_{1}d\bm{r}_{2}\dfrac{n_{\min}\left(\bm{r}_{1}\right)n_{\min}\left(\bm{r}_{2}\right)}{\left|\bm{r}_{1}-\bm{r}_{2}\right|}
\label{eq: hartree_s}
\end{equation}
where $n_{\text{min}}$ is the density constructed from  $\Psi_{\text{SD}}^{\text{min}}$. With the definitions of Eqs. (\ref{eq: kinetic}), (\ref{eq: ee-repul}), (\ref{eq: kinetic_s}), and (\ref{eq: hartree_s}), we define the exchange-correlation functional $E_{xc}[n^e]$ as 
\begin{equation}
E_{xc}\left[n^{e}\right]=T[n^{e}]-T_{s}[n^{e}]+V_{ee}[n^{e}]-J^{\prime}[n^{e}] \;\;\; .
\end{equation}
The Kohn-Sham energy minimization with the functional $E_{xc}\left[n^{e}\right]$ can be expressed as 
\begin{eqnarray}
E_{0} & = & \min_{n^{e}\rightarrow N}\left\{ G\left[n^{e}\right]+E_{xc}[n^e]+\int d\epsilon\; n^{e}\left(\epsilon\right)\upsilon^{e}\left(\epsilon\right)\right\} \notag \\
 & = & \min_{\Psi_{\text{SD}}}\left\{ \left\langle \Psi_{\text{SD}}\right|\hat{T}\left|\Psi_{\text{SD}}\right\rangle +J\left[n\right]+\right.   \notag \\
 &  & \left.+E_{xc}\left[n^{e}\right]+\int d\epsilon\; n^{e}\left(\epsilon\right)\upsilon\left(\epsilon\right)\right\} \;\;\; .
 \label{eq: KS_e}
 \end{eqnarray}
The minimization in the first equality in Eq.(\ref{eq: KS_e}) is performed within a space of $N$-rep. energy electron density, while it is replaced in the second equality by a search in the space of non-interacting wave function $\Psi_{\text{SD}}$. Since arbitrary $N$-rep. electron density can be generated from some $N$ orthonormal orbitals using Harriman's construction\cite{rf:harriman1980pra}, any $N$-rep. energy electron density defined in Eq.(\ref{eq:energy_dist}) is also given in terms of $\Psi_{\text{SD}}$. We note that part of kinetic and Hartree energies of the system is described with a non-interacting system in Eq.(\ref{eq: KS_e}). To derive the Kohn-Sham equation to achieve the minimization of Eq.(\ref{eq:energy_dist}) we formulate functional derivative of $E_{xc}[n^{e}]$ with respect to one-electron wave function $\varphi_i(\bm{r})$, thus, 
\begin{eqnarray}
\dfrac{\delta E_{xc}\left[n^{e}\right]}{\delta\varphi_{i}^{*}\left(\bm{r}\right)} & = & \int\int d\epsilon\; d\bm{r}^{\prime} \dfrac{\delta E_{xc}}{\delta n^{e}\left(\epsilon\right)}\dfrac{\delta n^{e}\left(\epsilon\right)}{\delta n\left(\bm{r}^{\prime}\right)}\dfrac{\delta n\left(\bm{r}^{\prime}\right)}{\delta\varphi_{i}^{*}\left(\bm{r}\right)}\delta\left(\bm{r}-\bm{r}^{\prime}\right)  \notag   \\
 & = & \int d\epsilon \; \dfrac{\delta E_{xc}}{\delta n^{e}\left(\epsilon\right)}\dfrac{\delta n^{e}\left(\epsilon\right)}{\delta n\left(\bm{r}\right)}\dfrac{\delta n\left(\bm{r}\right)}{\delta\varphi_{i}^{*}\left(\bm{r}\right)}    \notag    \\
 & = & \int d\epsilon \; \upsilon_{xc}[n^{e}](\epsilon)\delta(\epsilon-\upsilon(\bm{r}))\varphi_{i}(\bm{r})   \notag  \\
 & = & \left.\upsilon_{xc}[n^{e}](\epsilon)\right|_{\epsilon=\upsilon(\bm{r})} \varphi_{i}(\bm{r})  \;\;\; .   
 \label{eq: exc_deriv}
 \end{eqnarray}
where $\upsilon_{xc}[n^{e}](\epsilon)\equiv\delta E_{xc}[n^{e}]/\delta n^{e}\left(\epsilon\right)$ is the exchange-correlation potential for an electron on the energy coordinate $\epsilon$. We note that the definition of Eq.(\ref{eq:energy_dist}) is used to derive the third equality in Eq.(\ref{eq: exc_deriv}). It is  important to notice in Eq.(\ref{eq: exc_deriv}) that one-body potential which carries the effect of exchange and correlation among electrons is constant on the hyper surface of the same energy coordinate. As described above  this restriction would not be a serious disadvantage because of the postulate in Ref.\cite{rf:parr1981plenum}. Anyway, we have the Kohn-Sham equation  parallel to the conventional theory, thus, 
\begin{eqnarray}
\left[-\dfrac{1}{2}\nabla^{2}_{\bm{r}}+\upsilon_{\text{H}}(\bm{r})+\upsilon_{xc}(\epsilon)\right.   \notag   \\
\left.+\upsilon_{ext}(\bm{r})\right]\varphi_{i}(\bm{r}) & = & \eta_{i}\varphi_{i}(\bm{r})  \;\;\;  
\label{eq:Kohn-Sham}
\end{eqnarray}
where the first and second terms in the square bracket are, respectively, the kinetic and Hartree potential operators.\\
\indent We now examine the performance of the present approach using a prototype of the functional $E_{xc}[n^e]$. To this end we introduce the average electron density $\widetilde{n}^e (\epsilon)$ on the energy coordinate $\epsilon$ by dividing $n^e(\epsilon)$ with the spatial volume $\Omega(\epsilon)$ of the region with the energy coordinate $\epsilon$. Explicitly, $\widetilde{n}^e (\epsilon)$ is given by    
\begin{eqnarray}
\widetilde{n}^e\left(\epsilon\right) & = & n^e\left(\epsilon\right)/\Omega(\epsilon)  \label{eq: tilde_n} \\
\Omega(\epsilon) & = & \int d\bm{r}\delta\left(\epsilon-\upsilon(\bm{r})\right)    
\label{eq: omega}
\end{eqnarray}
An \textit{ad hoc} construction of the functional is to employ average density $\widetilde{n}^e (\epsilon)$ as a variable for some conventional density functional $E_{xc}[n]$. Then, the simplest exchange functional can be constructed in terms of the exchange energy of the homogeneous electron gas (HEG)\cite{rf:slater1951pr}, thus,  $E_{x}^{\text{HEG}}[n^{e}]=C_x\int d\epsilon\; n^e(\epsilon)\widetilde{n}^{e}(\epsilon)^{1/3}$ for each spin $(C_x=\tfrac{3}{4}(\tfrac{6}{\pi})^{1/3} )$.  We note that the functional is also dependent on the potential $\upsilon (\bm{r})$ of interest through Eqs. (\ref{eq: tilde_n}) and (\ref{eq: omega}).  Provided the average of the density gradient $\left|\nabla n(\bm{r})\right|_{\upsilon(\bm{r})=\epsilon}$ over the energy coordinate $\epsilon$ is given,  it is also possible to apply a generalized gradient approximations (GGA) to $E_{x}^{\text{HEG}}[n^{e}]$.  Below, we provide an exchange functional based on the Becke-Roussel (BRx) approach\cite{rf:becke1989pra, rf:takahashi2010jctc}. The BRx functional utilizes electron density of the hydrogenic atom as a model exchange hole for the real system.  Since it necessitates in its construction the second derivative $\nabla^2 n(\bm{r})$ of the electron density and the kinetic energy density $\tau (\bm{r})$, the information contents of the arguments to the functional is comparable to meta-GGA functionals. These derivatives are averaged over the energy coordinate and serve to construct the BRx functional $E_x^{\text{BR}}[n^e]$. We apply this functional combined with the LYP correlation functional\cite{rf:lee1988prb} to calculate the potential energy curve of an $\text{H}_2$ molecule around the bonding region.  

\begin{figure}
\scalebox{0.25}[0.25]{\includegraphics{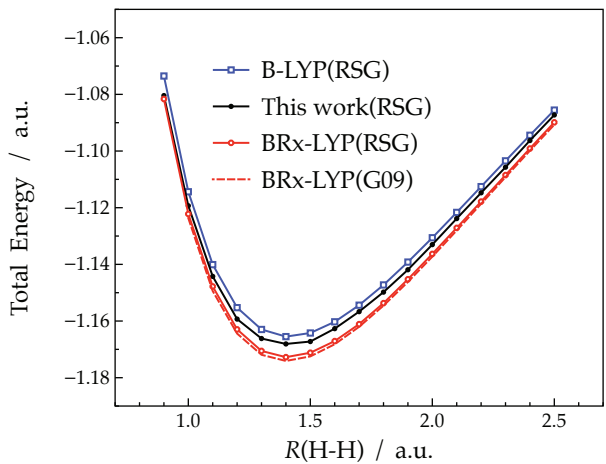}} % Here is how to import EPS art
\caption{\label{Fig2:epsart} Potential energy curves of $\text{H}_2$ molecule computed with the Kohn-Sham DFT utilizing real-space grid (RSG) method\cite{rf:chelikowsky1994prl, rf:takahashi2001jpca}. The exchange energy of this work was provided with the Becke-Roussel functional (BRx) using the energy electron density $n^e$, while the correlation energy was evaluated with the LYP formula as a functional of the electron density. The others were computed using functionals of the electron density. Result of BRx-LYP/aug-cc-pVTZ given by Gaussian 09 (G09) program suit is also shown as a reference. }
\end{figure}

The result is shown in Fig.\ref{Fig2:epsart}, where  it is also compared with the results of BRx-LYP and BLYP functionals\cite{rf:becke1988pra, rf:lee1988prb}
which use ordinary electron density $n$ as arguments. All of these calculations were performed using real-space grid (RSG) approach\cite{rf:chelikowsky1994prl, rf:takahashi2001jpca}. The electron density and wave functions for the BRx-LYP calculations were yielded by BLYP functional using ordinary electron density. To construct the energy electron density $n^e(\epsilon)$ with high resolution, the double grid technique was applied to interpolate the potential $\upsilon(\bm{r})$ on the neighboring original grids of RSG. Explicitly, five dense grid points were introduced on a coarse grid along each direction\cite{rf:ono1999prl}. It is realized in the figure that the potential energy of the present work which employs $n^e(\epsilon)$ gives the result reasonably close to the original BRx-LYP functional. The difference is smaller than  that of GGA functional (BLYP) from the original BRx-LYP. Thus, it is clearly demonstrated that the projection of the electron density $n(\bm{r})$ to the energy coordinate does not degrade the potential energy of $\text{H}_2$ seriously. We also found the similar trend for a hetero-diatomic molecule H-F with significant electronic polarization and for the double bond of $\text{C}_2 \text{H}_4$ molecule. It should be kept in mind, however, that the use of the pseudopotential leads to the shift of the exchange energy because the energy density for a core electron of a nucleus is simultaneously associated with the density of other nuclei.     \\
\indent We, next, consider the method to incorporate the static correlation into the  exchange functional. As illustrated in Fig.\ref{Fig1:epsart} the energy electron density $n_{\sigma}^{e}$ of the RKS solution at dissociation yields the same distribution as that of the isolated H atom provided the exact total electron density is given. Thereby, the  correct exchange energy can be realized with the energy density of the dissociated RKS wave function. Unfortunately, when we use the variable $\widetilde{n}^e(\epsilon)$ defined in Eqs. (\ref{eq: tilde_n}) and (\ref{eq: omega}) the dissociation curves of the UKS as well as RKS converge to the RKS energy produced with the conventional exchange functional. Since the volume $\Omega (\epsilon)$ at some coordinate $\epsilon$ of dissociated $\text{H}_2$ amounts to twice of the single atom, average electron density at $\epsilon$ becomes half of the corresponding density of the isolated atom. To solve this problem we introduce an intermediate state of which total electron density $n_0(\bm{r})$ is expressed in terms of the density $n_0^p$ of the isolated fragment $p$. Explicitly, $n_0(\bm{r})$ is given by
\begin{equation}
n_{0}\left(\bm{r}\right)=\sum_{p}n_{0}^{p}\left(\bm{r}\right)\;\;\;  .
\label{eq: n0}
\end{equation}
The choice of this reference state is motivated by the frozen density functional (FD) approach \cite{rf:wesolowski1993jpc} or the partition density functional theory (PDFT)  \cite{rf:elliott2010pra} where the total electron density of an interacting system is built from the densities of partitioned molecular regions. In the present work the density of Eq. (\ref{eq: n0}) serves to construct the exchange-correlation energy $E_{xc}[n_0]$ of the intermediate state. Then, the total exchange-correlation energy $E_{xc}$ of the density $n$ is expressed by sum of $E_{xc}[n_0]$ and the relaxation term $\Delta E_{xc}^{\text{relax}}$, thus, 
\begin{equation}
E_{xc}\left[n,n_{0}\right]=E_{xc}\left[n_{0}\right]+\Delta E_{xc}^{\text{relax}}\left[n,n_{0}\right] \;\;\; .
\label{eq: E_stat}
\end{equation}

\noindent The relaxation term describes the difference in the exchange-correlation energy between the states which correspond to the densities $n$ and $n_0$. In our approach  $\Delta E_{xc}^{\text{relax}}$ is evaluated by a functional of energy electron density, thus, $\Delta E_{xc}^{\text{relax}}=E_{xc}^{\upsilon}[n^e]-E_{xc}^{\upsilon}[n^e_0]$, where $n^e$ is derived from the spin-adapted wave function. At the dissociation limit the exact $n^e$ coincides with the fragment density $n_0^e$ by virtue of the property of the energy density as illustrated in Fig.\ref{Fig1:epsart}. Therefore, the relaxation term $\Delta E_{xc}^{\text{relax}}$ in Eq. (\ref{eq: E_stat})  completely vanishes  at the dissociation and only the energy $E_{xc}[n_0]$ remains, which guarantees the correct asymptotic behavior. In this approach the static correlation $E_{sc}$ is represented by the difference in the exchange correlation energy for the reference densities $n_0$ and $n_0^{e}$, thus, $E_{sc}=E_{xc}[n_0]-E_{xc}^{\upsilon}[n^e_0]$. The result for  $\text{H}_2$ dissociation is presented in Fig. 3, where we employ the Slater's LDA exchange functional $E_x^{\text{HEG}}$ that uses energy electron density $\widetilde{n}_0^e$ or $\widetilde{n}^e$  as argument.  The atomization energy is evaluated as $109.5$ kcal/mol which shows rather good agreement with the experimental value of 109 kcal/mol.  We emphasize that no adjustable parameter is introduced in the construction of the dissociation curves.  
A drawback in the method described above is that the density $n$ itself is determined independently of $E_{sc}$ because $E_{sc}$ employs only the reference density $n_0$ as an argument. To incorporate the effect of the static correlation in the variational calculation we next consider an implicit approach where the exchange potential $\upsilon_{x}$ of the interacting system is constructed through the linear response scheme with respect to the variation of the energy electron density $n(\epsilon)$. Explicitly, $\upsilon_{x}(\epsilon)$ is given as 
\begin{eqnarray}
\upsilon_{x}[n_{1}^{e}]\left(\epsilon\right) & = & \upsilon_{x}[n_{0}^{e}]\left(\epsilon\right)   \notag   \\
 & + & \int d\epsilon^{\prime}\left.\dfrac{\delta\upsilon_{x}[n^{e}]\left(\epsilon\right)}{\delta n^{e}\left(\epsilon^{\prime}\right)}\right|_{n^{e}=n_{0}^{e}}\delta n^{e}\left(\epsilon^{\prime}\right)
\label{eq:HNC}
 \end{eqnarray}
where $\delta n^e=n_1^{e}\left(\epsilon\right)-n_{0}^{e}\left(\epsilon\right)$ represents the density shift on the energy coordinate.
The existence of the exchange kernel ${\delta\upsilon_{x}\left(\epsilon\right)}/{\delta n\left(\epsilon^{\prime}\right)}$ is guaranteed by the one-to-one correspondence between the energy density and the external potential defined with some defining potential. It should be emphasized that we take an advantage of the property of the energy electron density in the construction of Eq.(\ref{eq:HNC}). That is, $\delta n^e$ can be an appropriate variable for the Taylor expansion since $n^e$  would be reasonably close to the distribution $n^e_0$ on the energy coordinate. Importantly, the exchange potential $\upsilon_x[n_1^e](\epsilon)$ as well as $n_1^e$ is optimized self-consistently  through  Eqs.(\ref{eq:Kohn-Sham}) and (\ref{eq:HNC}) under a given exchange kernel. In order to apply Eq.(\ref{eq:HNC}) to numerical calculations, we adopted the exchange kernel for HEG to Eq. (\ref{eq:HNC}), thus,     
\begin{equation}
\dfrac{\delta\upsilon_{xc}[n_{0}^{e}]\left(\epsilon\right)}{\delta n_{0}^{e}\left(\epsilon\right)}=\frac{4}{9}C_{x}\;\widetilde{n}_{0}^{e}\left(\epsilon\right)^{\frac{1}{3}}n_{0}^{e}\left(\epsilon\right)^{-1}
 \end{equation}
Correspondingly, the exchange energy $E_x[n_1^e]$ is given as
\begin{eqnarray}
E_{x}^{\text{HEG}}\left[n_{1}^{e}\right] & = & E_{x}^{\text{HEG}}\left[n_{0}\right]+   \notag \\
 &  & \int d\epsilon\;\delta n^{e}\left(\epsilon\right)\left\{ \frac{4}{3}C_{x}\widetilde{n}_{0}^{e}\left(\epsilon\right)^{\frac{1}{3}}+\right.     \notag    \\
 &  & \left.\;\;\;\;\;\;\;\;\;+\frac{2}{9}C_{x}\widetilde{n}_{0}^{e}\left(\epsilon\right)^{\frac{1}{3}}\dfrac{\delta n^{e}\left(\epsilon\right)}{n_{0}^{e}\left(\epsilon\right)}\right\}
 \label{eq:HNC_ene}      \;\; .
 \end{eqnarray}
The potential energy curve given by Eq. (\ref{eq:HNC_ene}) for $\text{H}_2$ dissociation is also plotted in Fig. 3.  
\begin{figure}[h]
\scalebox{0.28}[0.28]{\includegraphics{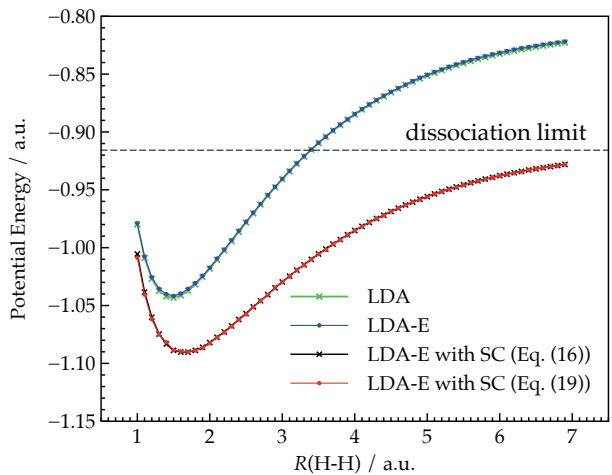}} % Here is how to import EPS art
\caption{\label{Fig3:epsart} $\text{H}_2$ dissociation curves calculated by spin restricted KS-DFT with the Slater's exchange functional  (LDA) and with the corresponding functional using energy electron density (LDA-E). The results which involve the static correlations (SC) employing Eqs. (\ref{eq: E_stat}) and (\ref{eq:HNC_ene}) are also presented. The value of the dissociation limit for the calculations with SC is evaluated as the energy at the interatomic distance of 14.0 a.u. All the calculations are performed with our real-space grid program code.  }
\end{figure}

It is worthy of note that the result obtained using Eqs.(\ref{eq:HNC})-(\ref{eq:HNC_ene}) and that given by  Eq.(\ref{eq: E_stat}) are hardly discernible, which implies the robustness of the linear-response approach represented with Eq. (\ref{eq:HNC}). We, thus, found that the variational approach can also be applied to the dissociation of a chemical bond. However, it should be noted that the potential curves with SC are erroneously stabilized in the middle range of the H-H distance ($R(\text{H-H}) = 3.0 - 4.0$ a.u.) and this  leads to the unfavorable elongation of the equilibrium bond distance. To avoid this problem will be the next issue in the calculation of SC using energy electron density $n^e(\epsilon)$.   \\
\indent In this Letter, we formulate the KS-DFT where the distribution $n^e(\epsilon)$ serves as a fundamental variable.  It was revealed that the KS-DFT based on the density $n^e(\epsilon)$ does not degrade seriously the quality of the exchange energy as compared to the conventional DFT. A property of crucial importance in the density $n(\epsilon)$ is that it can take into consideration the spatially non-local information of the electron density $n(\bm{r})$. By taking advantage of this property we developed a simple prototype of the exchange functional which offers the static correlation $E_{sc}$ in a bond dissociation. Although we found some unrealistic stabilization in the potential curve of $\text{H}_2$ at the middle range of the H-H distance, the dissociation energy is successfully evaluated showing a reasonable agreement with the experimental value. As a conclusion, the present approach with the density $n^e(\epsilon)$ can be potentially a new paradigm in DFT.  

\begin{acknowledgments}
The author wish to appreciate Professor N. Matubayasi in Osaka university for his helpful discussions, valuable comments and continuous encouragement to this work. The author is also grateful to A. Morita in Tohoku university for valuable suggestions. This paper was supported by the Grant-in-Aid for Scientific Research on Innovative Areas (No. 23118701) from the Ministry of Education, Culture, Sports, Science, and Technology (MEXT) and by the Grant-in-Aid for Challenging Exploratory Research (No. 25620004) from the Japan Society for the Promotion of Science (JSPS), and by the Nanoscience Program and the Computational Materials Science Initiative of
the Next-Generation Supercomputing Project.
\end{acknowledgments}

\bibliographystyle{apsrev4-1}
\bibliography{DFT}

\end{document}